\documentclass[lettersize,journal]{IEEEtran}

\usepackage{algorithmic}
\usepackage{array}
\usepackage{textcomp}
\usepackage{stfloats}
\usepackage{verbatim}
\usepackage{diagbox}
\usepackage{algorithm,amsbsy,amsmath,amsfonts,amssymb,epsfig,bbm,mathrsfs,multirow,amsthm,xcolor,cite}

\usepackage{epstopdf}
\usepackage{graphicx,caption,subcaption}
\newtheorem{definition}{Definition}
\newtheorem{proposition}{Proposition}
\usepackage{setspace}
\usepackage[hidelinks]{hyperref}
\usepackage[labelformat=simple]{subcaption}

\PassOptionsToPackage{hyphens}{url}
\usepackage{url}

\begin{document}

\title{Balancing Security and Accuracy: A Novel Federated Learning Approach for Cyberattack Detection in Blockchain Networks}


\author{Tran Viet Khoa, Mohammad Abu Alsheikh, Yibeltal Alem, and Dinh Thai Hoang	
        
\thanks{T.~V.~Khoa, M. Abu Alsheikh and Y. Alem are with the University of Canberra, Australia (e-mail: \{khoa.tran, mohammad.abualsheikh, yibe.alem\}@canberra.edu.au).} 

\thanks{D.~T.~Hoang are with the School of Electrical and Data Engineering, University of Technology Sydney, Australia (e-mail: hoang.dinh@uts.edu.au).}
}




\maketitle

\begin{abstract}

This paper presents a novel Collaborative Cyberattack Detection (CCD) system aimed at enhancing the security of blockchain-based data-sharing networks by addressing the complex challenges associated with noise addition in federated learning models. Leveraging the theoretical principles of differential privacy, our approach strategically integrates noise into trained sub-models before reconstructing the global model through transmission. We systematically explore the effects of various noise types, i.e., Gaussian, Laplace, and Moment Accountant, on key performance metrics, including attack detection accuracy, deep learning model convergence time, and the overall runtime of global model generation. Our findings reveal the intricate trade-offs between ensuring data privacy and maintaining system performance, offering valuable insights into optimizing these parameters for diverse CCD environments. Through extensive simulations, we provide actionable recommendations for achieving an optimal balance between data protection and system efficiency, contributing to the advancement of secure and reliable blockchain networks.


\end{abstract}


\section{Introduction}\label{sec:Int}
\IEEEPARstart{B}{lockchain} technology has witnessed rapid and transformative growth in recent years, revolutionizing various industries with its decentralized architecture. In a blockchain network, mining nodes independently store and manage their local data, organizing it into blocks. These nodes then undergo a rigorous validation process, such as proof of work~\cite{Bitcoin}, to select and add valid blocks to the global blockchain network. Once integrated, these blocks become immutable, unchangeable, and transparent to all participating nodes. With its core attributes of decentralization, immutability, transparency, and security, blockchain technology is finding widespread applications across numerous sectors, including finance, healthcare, supply chains, the Internet of Things (IoT), and smart grids~\cite{ali2018applications,da2021embedding, xie2019survey, cao2022blockchain}.

Blockchain networks are decentralized and thus face numerous cyber threats, such as denial of service (DoS)~\cite{Bitfinex}, flooding of transactions (FoT)~\cite{saad2020exploring}, brute pass (BP)~\cite{Top_hack}, and man-in-the-middle (MitM) attacks~\cite{wang2018attack}. These detrimental cyberattacks aim to disrupt blockchain operations, causing delays and enabling malicious activities, such as theft and system breaches. For example, a BP attack on Kucoin, a cryptocurrency in Singapore, caused a serious consequence with a significant loss of about~\$281 million~\cite{Top_hack}. In addition, an FoT attack on Bitcoin makes its memory full of dust transactions which leads to a significant delay on Bitcoin transactions costing about~\$700 million~\cite{saad2020exploring}. Moreover, a DoS attack targeted Bitfinex by sending a large number of synchronization (SYN) packages, forcing the platform to suspend operations for three hours~\cite{Bitfinex}. The rapid deployment of blockchains across vital sectors for managing sensitive data indicates that cyberattacks can directly impact and influence people's daily lives and well-being.


Collaborative cyberattack detection (CCD) systems demonstrate their efficiency in detecting various types of cyberattacks in blockchain networks with high accuracy~\cite{khoa2024collaborative, khoa2024collaborativelearningframeworkdetect}. However, these systems typically rely on federated learning, where workers must exchange raw trained models over the network. This exchange poses a significant security risk, as attackers can potentially reconstruct workers' local datasets by intercepting these raw models~\cite{wang2019beyond}. A natural solution to this problem is to protect the raw models during transmission by either adding noise to them or encrypting the data before sending it over the network. However, encryption methods are computationally intensive and require additional communication between the central server and workers to get the results~\cite{chang2023privacy}, especially in a blockchain network where minimizing computational time in processing blocks is crucial. 

In response to these challenges, previous studies~\cite{wei2020federated, zhou2022pflf, yang2021gain} have proposed enhancing the security of federated learning by adding noise to the trained sub-models before they are transmitted to a centralized server. This approach has been shown to effectively protect models from inference attacks. However, adding noise to trained models in federated learning-based cyberattack detection in blockchain networks presents several critical challenges. First, balancing the trade-off between privacy and model accuracy is difficult, as excessive noise can degrade detection performance. Second, the additional computational and communication overhead introduced by noise complicates the efficient operation of blockchain networks, which require rapid processing and minimal latency. Furthermore, identifying the right parameters for noise addition, such as the distribution, scale, and timing, requires careful tuning and may involve complex optimization problems, particularly in diverse and dynamic blockchain environments. To the best of our knowledge, no previous study has studied this problem.

In this paper, we propose a novel CCD system designed to enhance the security of blockchain-based data-sharing networks by addressing the critical challenges posed by noise addition in federated learning. Our approach integrates noise into the trained sub-models based on the theoretical foundations of differential privacy, before reconstructing the global model through transmission. We thoroughly investigate the impact of different noise types, i.e., Gaussian, Laplace, and Moment Accountant, on the accuracy of attack detection within the global model. This exploration is crucial, as the type and magnitude of noise significantly influence the detection accuracy, the convergence time of deep learning models in clusters, and the overall time required to generate the global model. By carefully balancing these factors, our work seeks to navigate the trade-offs between safeguarding against data leakage and maintaining system performance in terms of accuracy, convergence time, runtime, and the participation rate of clusters in the CCD model. Through extensive simulations, we evaluate the effectiveness of different noise parameters, offering concrete recommendations for achieving optimal results in various CCD environments and clustering configurations. The main contributions of this paper can be summarized as:


\begin{itemize}

    \item We propose a novel framework that enhances the security of CCD models in blockchain networks. In this framework, mining nodes can securely exchange trained models with robust protection against information leakage, ensuring the integrity and security of the data throughout the process.
    
    \item We introduce and systematically analyze the impact of various noise types, including Gaussian, Laplace, and Moment Accountant, within CCD models. Through comprehensive experimentation, we demonstrate the relationships between model accuracy, Differential Privacy (DP) parameters, and the number of participating workers, enabling us to identify optimal configurations for maximizing both security and performance.
    
    \item We rigorously evaluate the effects of noise addition on the overall runtime of CCD models, considering different noise types and cluster variations. Using real-world datasets, our simulations provide valuable insights into the performance of our proposed framework within the decentralized and resource-constrained environment of blockchain systems, highlighting its effectiveness and practicality.  
\end{itemize}

The rest of the paper is organized as follows. Section~\ref{sec:Related_Work} discusses related work and highlights key innovations of our work. Section~\ref{sec:proposed_CCD} introduces the proposed CCD in a blockchain-based data-sharing system. Section~\ref{sec:exp} presents our experiment setup, and then results are evaluated and discussed in detail in Section~\ref{sec:Performance_Evaluation}. Finally, Section~\ref{sec:conclusion} concludes the paper.

\section{Related work}
\label{sec:Related_Work}

\subsection{Machine Learning-Based Detection Methods for Blockchain Networks}

In~\cite{kim2021anomaly}, the authors propose an anomaly detection machine to detect anomalies (i.e., DoS and eclipse attacks) by analyzing the blockchain network traffic. The authors propose to use an autoencoder to detect anomalies (i.e., DoS and Eclipse attacks) from data instances. The experimental results show that their approach can achieve significant reductions in time complexity - up to 66.8\% during training and 85.7\% during testing - while maintaining robust system performance. In~\cite{liu2021lstm}, the authors propose to use condition generative adversarial networks (CGAN) to generate adversarial samples (i.e., Low-rate distributed DoS attacks) from normal traffic of both public and private datasets. They then use long short-term memory networks (LSTM) to learn the relationship between samples to detect attacks. The simulation results show that with various machine learning and deep learning classifiers, the precisions can reach up to 95\%. In~\cite{cao2021blockchain}, the authors propose to use blockchain technology to detect link flooding attacks, a new type of distributed DoS (DDoS) in blockchain networks. The authors used traceroute records to analyze and detect the attack. The simulation results show that their approach can achieve a detection rate of nearly 100\%. 

It can be observed that the approaches mentioned above depend on centralized algorithms, which require the aggregation of all network data onto a central server for analysis. However, blockchain technology is fundamentally decentralized, making it inherently difficult, if not impractical, to consolidate data from all nodes into a single central server for processing. Moreover, centralizing data collection introduces significant risks, including potential privacy breaches and the vulnerability of a single point of failure. As a result, these centralized approaches are neither effective nor secure for deployment in blockchain systems.

\subsection{Collaborative Detection Methods for Blockchain Networks}

In~\cite{alkadi2020deep}, the authors propose a deep learning-based collaborative intrusion detection to protect cloud and IoT networks. The authors employ a bidirectional long short-term memory (BiLSTM) deep learning technique to develop intrusion detection and blockchain to protect the privacy of systems. The simulation results with two ubiquitous datasets show that their framework can reach nearly 100\% in detecting anomalies. In~\cite{liang2021data}, the authors propose a data fusion approach for collaborative intrusion detection in a blockchain-based system to detect anomalies in a blockchain system. The authors design a data fusion model to analyze data groups (e.g., the KDD - a computer network cyberattack dataset) in blockchain networks. The simulation results show that their approach achieves an impressive True Positive Rate of up to 97\% in detecting anomalies within blockchain-based systems. While the methods proposed in~\cite{alkadi2020deep} and~\cite{liang2021data} may exhibit high accuracy in anomaly detection, they fall short in identifying and distinguishing the specific types of attacks concealed within those anomalies. 


In~\cite{khoa2024collaborative}, the authors propose a collaborative learning model that can detect cyberattacks in blockchain networks. The authors first build a blockchain network attack traffic (BNaT) dataset that includes the normal and attack behaviors of a blockchain network. They then propose a collaborative learning model based on federated learning to detect different types of attacks. The simulation and real-time experimental results show that their approach can achieve 98.6\% accuracy in detecting attacks. In~\cite{khoa2024collaborativelearningframeworkdetect}, the authors propose a collaborative learning model for attack detection in smart contracts and transactions in a blockchain system. The authors collected a dataset of normal and attack behaviors in smart contracts and transactions. After that, they propose to convert the important features of transactions and smart contracts into images and then use a deep convolutional neural network (CNN) to classify attacks. The simulation and real-time experimental results show that their model can achieve an accuracy of 94\% in detecting attacks in transactions and smart contracts. While the approaches in~\cite{khoa2024collaborative} and~\cite{khoa2024collaborativelearningframeworkdetect} leverage collaborative learning models to detect attacks in decentralized blockchain networks, their reliance on transferring raw gradients among networks introduces significant vulnerabilities. This highlights the distinct advantage of the proposed model in offering a more secure and accurate detection mechanism.


In~\cite{geiping2020inverting}, the authors reveal that an inverting process can effectively reconstruct original data from raw gradients. Their experimental results clearly show that exchanging raw gradients in collaborative learning methods, such as federated learning, fails to secure the privacy of data within worker networks. Consequently, sensitive information remains vulnerable to exposure by unauthorized parties who can analyze parameter variations in the trained models, as evidenced in~\cite{melis2019exploiting, wang2019beyond, zhu2019deep}. This underscores a critical weakness in privacy in these methods.
Therefore, in this work, unlike earlier approaches, which often rely on centralized systems or partial decentralization, we propose a fully decentralized CCD solution that uses differential privacy to train a data-leaking protection CCD model on geographically distributed CCD devices. Our solution adds a layer of data protection to individual CCD clusters, preventing any specific user's data from being exposed or reconstructed and maintaining data integrity and security throughout the entire process. 

\section{Proposed CCD in Blockchain-Based Data Sharing}
\label{sec:proposed_CCD}

\subsection{The Fundamental of Blockchain Networks}
\label{sec:model}

\begin{figure*}[!t]
    \centering
    \includegraphics[width=\linewidth]{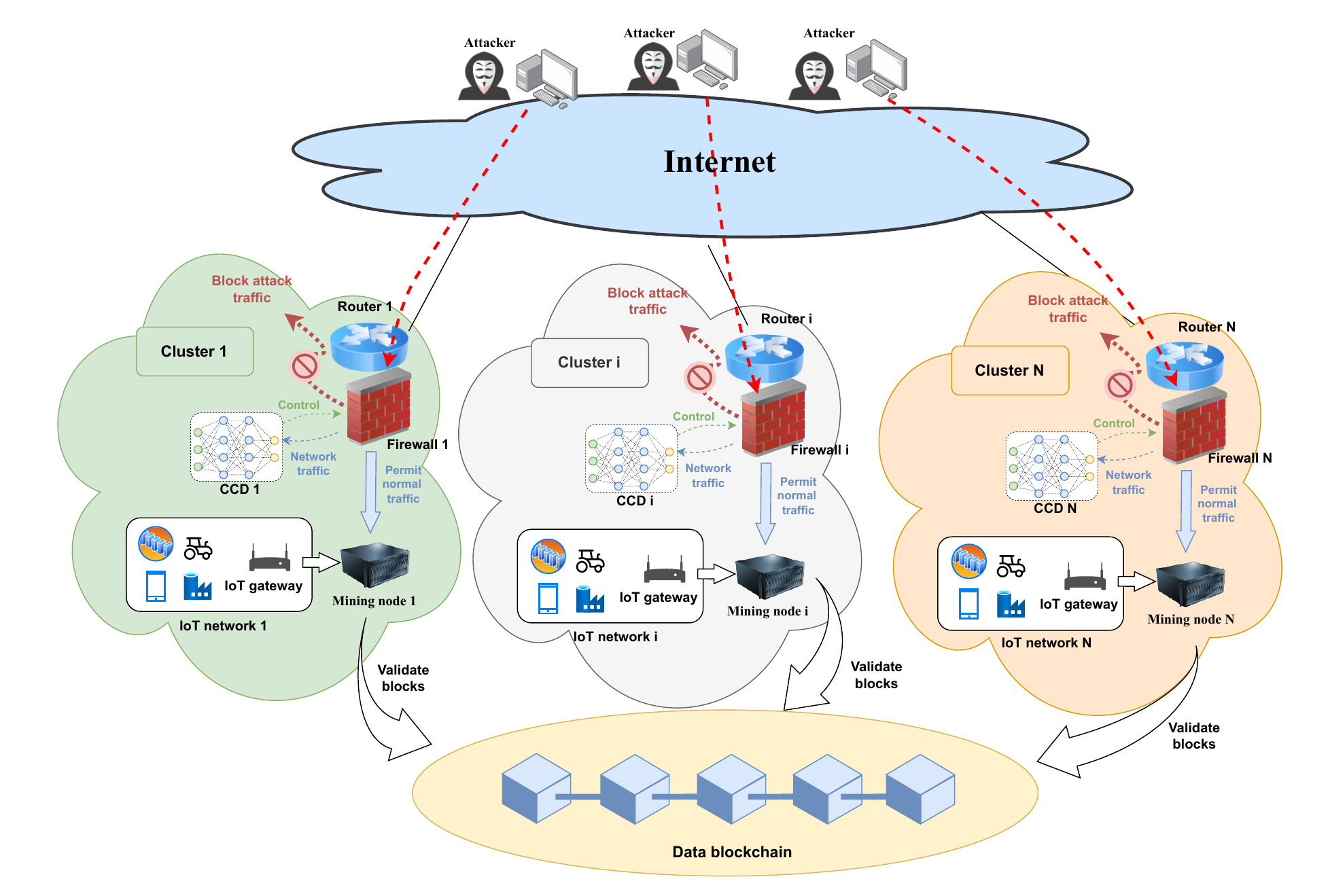}
    \caption{The proposed CCD systems deployed in a blockchain-based data-sharing network for collaborative training-process and real-time cyberattack detection.}
    \label{fig:Collaborative_sys_model}
    \vspace*{-0.5cm}
\end{figure*}

Blockchain technology represents a transformative approach to data sharing among individuals, organizations, and governments, enabling secure and decentralized transactions without needing third-party validation. In a blockchain-based data-sharing system, multiple mining nodes collaboratively store and process data, primarily consisting of transactions generated from network traffic in an IoT environment. These transactions are aggregated into blocks, upon which the mining nodes execute the mining process - a crucial function that employs sophisticated cryptographic algorithms. This process is vital for achieving consensus across all nodes, ensuring the blockchain remains consistent, tamper-resistant, and trustworthy. Once a block is validated through the consensus mechanism, it is permanently added to the blockchain, creating an unalterable and transparent record of all transactions. Blockchain-based data sharing offers significant advantages over traditional database systems, including eliminating central control, enhanced transparency, flexible data access, ensured data integrity and immutability, scalability, and improved traceability. These attributes make blockchain a superior choice for secure and efficient data management.


\subsection{The Proposed System Model}

Fig.~\ref{fig:Collaborative_sys_model} describes our proposed CCD system in a blockchain-based data-sharing network. In this model, we have $N$ data-sharing clusters, each including an IoT network for data collection, a CCD device for cyberattack detection, and a mining device for blockchain operations.
\begin{itemize}
    \item \textbf{IoT network}: Each IoT network collects data from multiple sensing nodes distributed across a geographical area. The data serves a variety of critical applications, such as smart cities, healthcare, transportation, agriculture, and environmental monitoring. The IoT gateway plays a pivotal role in this process by aggregating data from all connected IoT devices and converting it into transactions ready for deployment onto a blockchain system. To ensure data authenticity and ownership, the IoT gateway also appends its digital signature to each data transaction, providing a secure and verifiable layer of authentication.
    
    \item \textbf{Mining node}: The data transactions are sent to the mining node in each data-sharing cluster. The mining device performs the critical role of receiving, validating, and adding transactions to the global blockchain network. The mining nodes perform a mining process using a consensus protocol to validate the data, e.g., proof-of-work~\cite{Bitcoin}. The validated data of a mining node that passes over the mining process will be put into a global blockchain network. After that, the data is immutable and transparent to mining nodes in all clusters.
    
    \item  \textbf{CCD device}: The CCD device initially leverages its locally received transaction data to train its CCD models. These models function as a critical ``filter'', detecting and intercepting potential attacks before any data is permitted into the cluster's network. Moreover, all CCD devices are interconnected via peer-to-peer (P2P) communication, eliminating the need for an intermediary central server or authority, thus enhancing the system's security and decentralization.

\end{itemize}



\begin{figure*}[!t]
    \centering
    \includegraphics[width=.8\linewidth, trim=1cm 0.5cm 1cm 0cm]{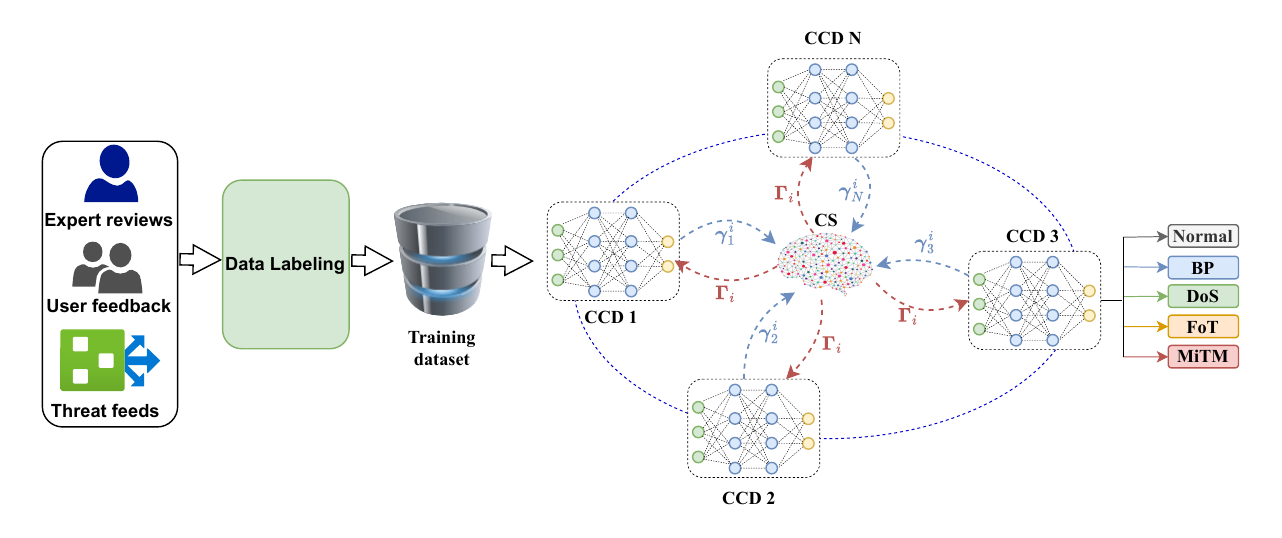}
    \caption{The collaborative learning process among clusters.}
    \label{fig:CCDS_system}
    \vspace*{-0.5cm}
\end{figure*}

The CCD model is trained with a training dataset that includes various types of attacks within a blockchain network. The training dataset can be updated regularly using the collected network data of a blockchain network. Training data can be labeled using expert reviews, user feedback schemes, and automated threat feeds, e.g., honeypots, security alerts, deceptive environments, and cryptographic verification. The real-time collection of CCD data is depicted in Fig.~\ref{fig:CCDS_system}.

We denote $\boldsymbol{D}_{n}^{i}$ as the dataset of cluster $n$, with $n \in \{0,...,N\}$ at iteration $i$. Besides, each cluster also has a deep learning model that serves as the CCD model. The output of the deep learning model in a cluster $n$ at iteration $i$ can be calculated as follows~\cite{goodfellow}:
\begin{equation}
\begin{aligned}
\label{eqn1}
\boldsymbol{I}_{n}^{i} = \boldsymbol{H}_n\boldsymbol{D}_{n}^{i},
\end{aligned}
\end{equation}
where $\boldsymbol{I}_{n}^{i}$ is the output prediction at iteration $i$ of the CCD model in cluster $n$ and $\boldsymbol{H}_n$ is a transfer function of the CCD model in cluster $n$.

We denote $\boldsymbol{Y}_n$ as the vector of labels corresponding to $\boldsymbol{D}_{n}$ and $\mathcal{L}(\cdot)$ as the loss function of the CCD model. The gradient (trained model) $\nabla\boldsymbol{\theta}_{n}^{i}$ of the deep model at cluster $n$ at iteration $i$ is denoted as $\nabla\mathcal{L}(\boldsymbol{\theta}_{n}^{i};\boldsymbol{D}_{n}^{i})$. We can then update the weights of the CCD model as follows:
\begin{equation}
\begin{aligned}
\label{eqn3}
\boldsymbol{\theta}_{n}^{i} = \boldsymbol{\theta}_{n}^{i-1} - \mu\nabla_{\boldsymbol{\theta}_{n}}\mathcal{L}(\boldsymbol{\theta}_{n};\boldsymbol{D}_{n}^{i}),
\end{aligned}
\end{equation}
where $\mu$ is the learning rate and $\nabla_{\boldsymbol{\theta}_{n}}\mathcal{L}(\boldsymbol{\theta}_{n};\boldsymbol{D}_{n}^{i})$ is the gradient of $\mathcal{L}(\cdot)$ with respect to $\boldsymbol{\theta}_{n}$ computed for $\boldsymbol{D}_{n}^{i}$. The local updates $\bigtriangleup\boldsymbol{\theta}_{n}^{i}=\boldsymbol{\theta}_{n}^{i}-\boldsymbol{\theta}_{n}^{i-1}$ are clipped to a maximum norm $\vartheta$ as follows:
\begin{equation}
\bigtriangleup\boldsymbol{\theta}_{n}^{i}\leftarrow\bigtriangleup\boldsymbol{\theta}_{n}^{i}\min\left(1,\frac{\vartheta}{\Vert\bigtriangleup\boldsymbol{\theta}_{n}^{i}\Vert}\right).
\end{equation}

The clipping bounds the influence of any single update and hence stabilizes the learning process~\cite{abadi2016deep}.



\subsection{Privacy-preserving CCD}

Each cluster $j$ maintains a CCD sub-model $\textbf{w}_j$, $j \in \{1,\ldots,N\}$ in the CCD device. The CCD adds different types of noises to its local sub-model. These noises will protect trained models while they are exposed over the network. We denote $(\epsilon,\delta)$ as a couple of noise parameters, $\boldsymbol{D}$ and $\boldsymbol{D'}$ as two adjacent datasets, i.e., differing by one data element. In this scenario, $\epsilon$ defines the distinguishable limit of outputs for $\boldsymbol{D}$ and $\boldsymbol{D'}$. Besides, the parameter $\delta$ indicates the probability that the ratio of the probabilities for $\boldsymbol{D}$ and $\boldsymbol{D'}$ exceeds $e^{\epsilon}$ in the added noise model. $\epsilon$ and $\delta$ are non-negative parameters that measure the privacy loss-$\epsilon$ is the privacy budget and $\delta$ is the probability of the privacy loss exceeding $\epsilon$. For any given $\delta$, an added noise with a higher $\epsilon$ increases the clarity with which adjacent datasets can be distinguished, thus raising the risk of privacy breaches~\cite{wei2020federated}. We can define an added noise to the CCD model as follows:

\begin{definition}$(\epsilon,\sigma)$-differential privacy~\cite{dwork2014algorithmic}: A transfer function $\mathcal{R}\!\!: \mathcal{D} \rightarrow \mathcal{O}$ with $\mathcal{D}$ as input domain and range $\mathcal{O}$ satisfies the $(\epsilon,\sigma)$-differential privacy, for all subsets of outputs $ \mathcal{I} \subseteq \mathcal{O}$ of two adjacent datasets $\boldsymbol{D}, \boldsymbol{D'} \in \mathcal{D}$,
\begin{equation}
\begin{aligned}
\label{eqn0}
Pr[\mathcal{R}(\boldsymbol{D}) \in \mathcal{I}] \leq \exp{(\epsilon)} Pr[\mathcal{R}(\boldsymbol{D'}) \in \mathcal{I}] + \delta.
\end{aligned}
\end{equation}
\end{definition} 

We denote $t$ as a real-valued function using a differential security mechanism that involves adding noise. This noise is adjusted according to the function's sensitivity $\Delta t$. $\Delta t$ is defined as the maximum change in the function’s output when an individual entry in the dataset is modified as follows:
\begin{equation}
\bigtriangleup t=\max_{\boldsymbol{D},\boldsymbol{D'}}\left\Vert t(\boldsymbol{D})-t(\boldsymbol{D'})\right\Vert. 
\end{equation}

In this paper, we study various types of noises, including Gaussian, Laplace, and Moment Accountant, to evaluate their effects on the accuracy and the number of clusters in the data-sharing network. In our proposed model, each cluster has its private dataset. 
At this stage, the CCD model in cluster $n$ has to send $\boldsymbol{\theta}_{n}^{i}$ to perform collaborative learning. To mitigate the risk of exposing raw the trained models of the neural networks over the transmission channel. We add noises to the raw local CCD model to meet the requirements for $(\epsilon,\sigma)$-differential privacy. We denote $\alpha_{n}^{i}$ as the added noise to the trained model of cluster $n$ at iteration $i$ as follows:
\begin{equation}
\begin{aligned}
\label{eqn4}
\boldsymbol{\gamma}_{n}^{i} = \boldsymbol{\theta}_{n}^{i} + \alpha_{n}^{i}.
\end{aligned}
\end{equation}

The amount of noise is computed according to the sensitivity of the function and the privacy level $\epsilon$. Furthermore, in this work, we use $(\epsilon,\delta)$ as a couple of noise parameters to protect the raw model in cluster $n$ over the network. Different noise mechanisms can be used in $\alpha_{n}^{i}$, such as Gaussian, Laplace, and Moment Accountant Gaussian. We denote $x$ as the variable representing the noise value, and $c$ as the standard deviation. The details are described as follows:

\subsubsection{Gaussian Mechanism}
The Gaussian mechanism has the probability density function (PDF) as follows:
\begin{equation}
\begin{aligned}
\label{eqn5}
f_G(x|c) = \frac{1}{c\sqrt{2\pi}} \exp\left(-\frac{x^2}{2c^2}\right).
\end{aligned}
\end{equation}

We denote $|\boldsymbol{D}_{n}|$ as the size of the dataset of the cluster $n$. With $\vartheta$ as the clipping threshold for the bounding of $\boldsymbol{\theta}_{n}^{i}$, the sensitivity $\Delta t$ of function $t$ can be defined as follows~\cite{wei2021user}:
\begin{equation}
\begin{aligned}
\label{eqn6}
\Delta t=\frac{2 \mu \vartheta}{|\boldsymbol{D}_{n}|}.
\end{aligned}
\end{equation}

We then can use equation~(\ref{eqn6}) to calculate the standard deviation of Gaussian noise~\cite{wei2020federated}: 
\begin{equation}
\begin{aligned}
\label{eqn7}
c &= \frac{\Delta t \sqrt{2 \ln(1.25/\delta)}}{\epsilon}
  &= \frac{e_1 \sqrt{2 \ln(1.25/\delta)}}{|\boldsymbol{D}_{n}|\epsilon},
\end{aligned}
\end{equation}
where $e_1$ is the constant number of Gaussian distribution. 

\subsubsection{Laplace Mechanism}
The PDF of the Laplace distribution is described as follows~\cite{yang2023csra}:
\begin{equation}
\begin{aligned}
\label{eqn8}
f_L(x|b) = \frac{1}{2b} \exp\left(-\frac{|x|}{b}\right),
\end{aligned}
\end{equation}
where $b$ is the scale parameter. According to~\cite{yang2023csra} and using equation~(\ref{eqn6}), we can calculate $b$ as follows: 
\begin{equation}
\begin{aligned}
\label{eqn9}
b = \frac{\Delta t}{\epsilon} = \frac{e_2 \vartheta}{|\boldsymbol{D}_{n}| \epsilon},
\end{aligned}
\end{equation}
where $e_2$ is the constant number of Laplace distribution.

\subsubsection{Moment Accountant Mechanism}
The Moment Accountant is proposed in~\cite{abadi2016deep}. Although this mechanism uses the Gaussian distribution as in equation~(\ref{eqn5}), it uses another formula to calculate $c$ to strengthen the composition theorem. According to~\cite{wei2021user}, we denote $q$ as the batch size and $T$ as the maximum number of iterations. The standard deviation of noise in Federated Learning can be calculated as follows: 
\begin{equation}
\begin{aligned}
\label{eqn10}
c &= \frac{\Delta t \sqrt{2 q T \ln(1/\delta)}}{\epsilon}
  &= \frac{e_3 \sqrt{2 q T\ln(1/\delta)}}{|\boldsymbol{D}_{n}|\epsilon},
\end{aligned}
\end{equation}
where $e_3$ is the constant number of the Momentum Accountant.

\subsection{Aggegation Process}
As shown in Fig.~\ref{fig:CCDS_system}, after adding noise as in equation~(\ref{eqn4}), the trained models of all clusters are sent to a centralized server (CS) for aggregation. This server can either be a standalone server or integrated into the CCD of a cluster. The CS then consolidates all the aggregated parameters to generate a new global model $\boldsymbol{\Gamma}_i$ at iteration $i$ as follows:
\begin{equation}
\begin{aligned}
\label{eqn11}
\boldsymbol{\Gamma}_{i} = \frac{1}{N} \sum_{n=1}^N \boldsymbol{\gamma}_{n}^{i}.
\end{aligned}
\end{equation}

The global model $\boldsymbol{\Gamma}_{i}$ is then used to update the weight matrices of the CCD models of all clusters before starting the next learning iteration. This process continues with the increase of $i$ until the $i$ approaches the maximum number of iteration $T$. When $i=T$, the optimal global model $\boldsymbol{\Gamma}_{opt}$ will be created. The CCD devices of all clusters will use $\boldsymbol{\Gamma}_{opt}$ to analyze and predict attacks in their local networks. The summary of the system is described in Algorithm~\ref{al:DP_FL}.

From equations~(\ref{eqn5}),~(\ref{eqn7}),~(\ref{eqn8}), and~(\ref{eqn9}) we can see that the PDF of Gaussian distribution and Laplace are proportional to the size of the local dataset of the cluster. Besides, we can see in equations~(\ref{eqn10}) that the Gaussian distribution varies based on both the size of the local dataset of cluster $n$ and the number of running iterations. The variations of distributions cause various impacts on the final global model as well as the accuracy of attack detection. More details can be discussed below.

\begin{algorithm}[t]
	\algsetup{linenosize=\tiny}
	\caption{The securing collaborative learning algorithm}
	\label{al:DP_FL}
	\begin{algorithmic}[1]
	    \WHILE{$i \leq T$}
	        \FOR{$\forall n \in N$}
                \STATE \textit{The DL model of cluter $n$ performs}:
        		\STATE Learn $\boldsymbol{D}_{n}^{i}$ to create the prediction output $\boldsymbol{I}_{n}^{i}$. 
        		\STATE Calculate the gradient $\nabla\boldsymbol{\theta}_{n}^{i}$.
        		\STATE Update its parameters $\boldsymbol{\theta}_{n}^{i}$.
                \STATE Add noise $\alpha_{n}^{i}$ into $\boldsymbol{\theta}_{n}^{i}$ to create $\boldsymbol{\gamma}_{n}^{i}$.
                \STATE Send $\boldsymbol{\gamma}_{n}^{i}$ to CS.
		    \ENDFOR
            \STATE \textit{The centralized server performs:}
		    \STATE Calculates the global model $\boldsymbol{\Gamma}_{i}$.
		    \STATE {$i=i+1$.}
		    \STATE \textit{The DL models in all clusters perform:} 
            \STATE Use $\boldsymbol{\Gamma}_{i}$ to update their local models. 
		\ENDWHILE
        \STATE \textit{The centralized server performs:}
        \STATE Generates the optimal global model $\boldsymbol{\Gamma}_{opt}$.
        \STATE Sends $\boldsymbol{\Gamma}_{opt}$ to all clusters.
	    \STATE \textit{The clusters perform:} 
        \STATE Use $\boldsymbol{\Gamma}_{opt}$ to generate the prediction output from the input data of clusters. 
 	\end{algorithmic}
\end{algorithm}


\begin{proposition}
The aggregation of a collaborative learning in a blockchain network is ($\bar{\epsilon},\bar{\delta}$)-differentially private, where $\bar{\epsilon}$ and $\bar{\delta}$ are bounded as follows:


\begin{equation}
\bar{\epsilon}\leq\epsilon\sqrt{2NT\ln\left(\frac{1}{\bar{\delta}}\right)}+NT\epsilon\left(\exp(\epsilon)-1\right),
\end{equation}

\begin{equation}
\bar{\delta}\leq NT\delta+\grave{\delta},
\end{equation}
where $\grave{\delta}\geq0$ helps bound the cumulative privacy loss over multiple aggregations (multiple iterations and devices), $T$ is the number of training iterations, $N$ is the number of clusters, and $\epsilon$ and $\delta$ are the privacy cost of each CCD device.

\end{proposition}

\begin{proof}

To prove this proposition, we can define the bound of the privacy cost in blockchain networks by applying parallel composition~\cite{mcsherry2009privacy} and sequential composition~\cite{dwork2006our}. In particular, the privacy bound should account for both the best-case scenario of parallel composition, where the CCD datasets of clusters are independent, and the worst-case scenario of sequential composition, where a data sample can appear in multiple CCD clusters.

In the best-case scenario of parallel composition, i.e.,~the CCD devices learn on independent datasets (disjoint chunks of attack data),  the overall privacy budget remains the same as individual privacy cost~\cite{mcsherry2009privacy}. Thus, $(\bar{\epsilon},\bar{\delta})= (\epsilon,\delta)$.

In the worst-case scenario of sequential composition, the privacy cost of the collaborative learning model can be calculated as $(\bar{\epsilon},\bar{\delta})= (NT\epsilon,NT\delta)$, for $N$ clusters and $T$ training iterations. But, we can still apply the advanced composition theorem~\cite{dwork2010boosting} to find a tighter bound.

\begin{equation}
\bar{\epsilon}=\epsilon\sqrt{2NT\ln\left(\frac{1}{\bar{\delta}}\right)}+NT\epsilon\left(\exp(\epsilon)-1\right),
\end{equation}

\begin{equation}
\bar{\delta}= NT\delta+\grave{\delta}.
\end{equation}

The proof is now completed. 
\end{proof}

Proposition 1 that establishes the aggregation of trained models with added noise as ($\bar{\epsilon},\bar{\delta}$)-differential private is crucial in federated learning for several reasons. Firstly, it ensures the protection of individual data privacy by mathematically guaranteeing that the inclusion of noise limits the probability of identifying specific data points within the shared model parameters. This protection is particularly important in our system where sensitive information is distributed across multiple mining nodes, as in federated learning, where each mining node operates on local data. Secondly, this proposition strikes a critical balance between privacy and utility. While noise is essential for maintaining privacy, it can also degrade the performance of the global model. The bounds defined by ($\bar{\epsilon},\bar{\delta}$)-differential privacy ensure that enough noise is added to protect privacy without significantly diminishing the model's accuracy and effectiveness. Thus, this proposition is vital for designing federated learning systems that are both secure and efficient, enabling the practical application of federated learning in our proposed framework.


\section{Experiment Setup and Evaluation Method}\label{sec:exp}

In this section, we present the datasets we used in the experiment and provide details of the evaluation method we used to evaluate the systems' performance.

\subsection{Blockchain Cyberattack Dataset}
In this paper, we use the Blockchain Network Attack Traffic dataset~\cite{khoa2024collaborative}. This dataset was created based on various experiments on a private Ethereum network. The dataset includes both normal network traffic and four common network traffic attacks, i.e., denial of service (DoS), flooding of transactions (FoT), brute pass (BP), and man-in-the-middle (MitM) attacks. They are ubiquitous attacks that have caused many serious consequences in blockchain networks for years.

\begin{table}[!t]
\centering
\caption{Features of the dataset.}
\label{tab:features}
\resizebox{0.49\textwidth}{!}{
\begin{tabular}{|l|l|l|l|} 
\hline
\multicolumn{1}{|c|}{\begin{tabular}[c]{@{}c@{}}\#\\\end{tabular}} & \multicolumn{1}{c|}{\textbf{Features name}} & \multicolumn{1}{c|}{\textbf{Description}} \\ 
\hline
1 & \textit{duration} & \begin{tabular}[c]{@{}l@{}} duration of the connection\end{tabular} \\ 
\hline
2 & \textit{protocol\_type} & \begin{tabular}[c]{@{}l@{}}protocol type (e.g., tcp, udp)\end{tabular} \\ 
\hline
3 & \textit{service} & \begin{tabular}[c]{@{}l@{}}type of service (e.g., http, https)\end{tabular} \\ 
\hline
4 & \textit{src\_bytes} & \begin{tabular}[c]{@{}l@{}} data size from source\end{tabular} \\ 
\hline
5 & \textit{dst\_bytes} & \begin{tabular}[c]{@{}l@{}}data size from destination\end{tabular} \\ 
\hline
6 & \textit{flag} & \begin{tabular}[c]{@{}l@{}}status of the connection \\ (i.e., normal and error)\end{tabular} \\ 
\hline
7 & \textit{count} & \begin{tabular}[c]{@{}l@{}} number of connections has \\ the same source IP\end{tabular} \\ 
\hline
8 & \textit{srv\_count} & \begin{tabular}[c]{@{}l@{}} number of connections has \\ the same source service\end{tabular} \\ 
\hline
9 & \textit{serror\_rate}& \% of connections with `SYN' errors \\ 
\hline
10 & \textit{same\_srv\_rate} & \% of connections with same service\\ 
\hline
11 & \textit{diff\_srv\_rate} & \begin{tabular}[c]{@{}l@{}}\% of connections with~different \\services~ ~\end{tabular} \\ 
\hline
12 & \textit{srv\_serror\_rate} & \% of connections with `SYN' errors\\ 
\hline
13 & \textit{srv\_diff\_host\_rate}  & \% of connections to different host \\ 
\hline
14 & \textit{dst\_host\_count}  & \begin{tabular}[c]{@{}l@{}} number of connections has \\ the same destination IP\end{tabular} \\ 
\hline
15 & \textit{dst\_host\_srv\_count}  & \begin{tabular}[c]{@{}l@{}} number of connections has \\ the same destination service\end{tabular} \\ 
\hline
16 & \textit{dst\_host\_same\_srv\_rate} & \begin{tabular}[c]{@{}l@{}}\% of connections with the same \\target host and service\end{tabular} \\ 
\hline
17 & \textit{dst\_host\_diff\_srv\_rate}  & \begin{tabular}[c]{@{}l@{}}\% of connections with the same \\target host but different services\end{tabular} \\ 
\hline
18 & \textit{dst\_host\_same\_src\_port\_rate} & \begin{tabular}[c]{@{}l@{}}\% of connections with the same \\target host and source port\end{tabular} \\ 
\hline
19 & \textit{dst\_host\_serror\_rate}  & \% of connections with `SYN' errors \\ 
\hline
20 & \textit{dst\_host\_srv\_diff\_host\_rate} & \begin{tabular}[c]{@{}l@{}}\% of connections with the same \\target host and service but different \\source hosts\end{tabular} \\ 
\hline
21 & \textit{dst\_host\_srv\_serror\_rate} & \% of connections with `SYN' errors~ ~ \\
\hline
\end{tabular}
}
\end{table}

\begin{itemize}
    \item \textbf{DoS attacks} aim to disturb the working of blockchain networks by preventing legitimate users from accessing the networks. DoS attacks involve sending an excessive amount of traffic to overwhelm a blockchain network. As a result, DoS attacks can degrade the network availability and response time.
    \item \textbf{FoT attacks} work by flooding blockchain networks with numerous ineffective transaction requests. This attack can create a large number of pending transactions to delay the consensus protocol and mining process.
    \item \textbf{BP attacks} involve guessing passwords and encrypted keys by trying combinations and random guesses. BP attackers aim to gain unauthorized access to protected resources in a blockchain network.
    \item \textbf{MitM attacks} work when an attacker intercepts and potentially alters the data exchanged between two entities in a blockchain network. MitM attackers generally target blockchain networks to gain sensitive information, such as private keys or transaction details.
\end{itemize}

The dataset comprises 210,000 samples, including 147,000 for training and 63,000 for testing. The dataset was collected from three distinct Ethereum mining nodes, providing a diverse and comprehensive set of data points. The data samples are categorized into five classes: one representing normal behaviour and four types of attacks (DoS, FoT, BP, and MitM attacks). The dataset includes 21 key features extracted from the blockchain network traffic as described in Table~\ref{tab:features}. The features are extracted from raw network traffic to distinguish normal behaviour from attacks.

\begin{figure}[!t]
    \centering
    \begin{subfigure}{\linewidth}
        \includegraphics[width=1.0\linewidth]{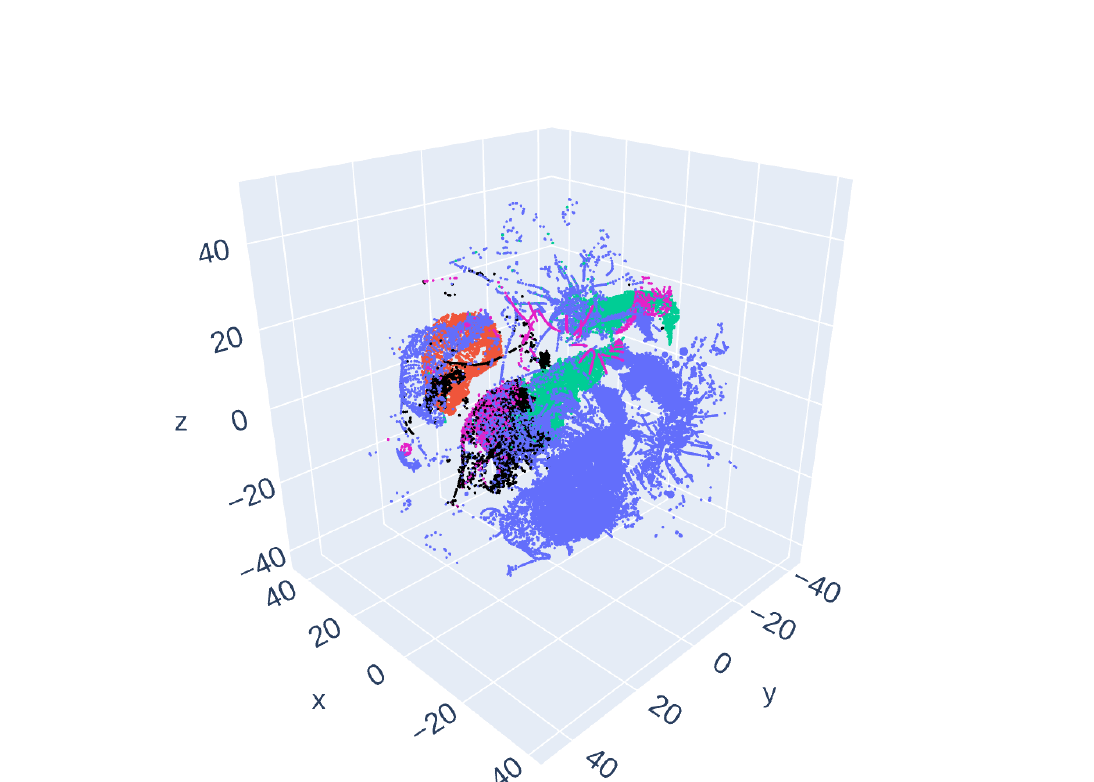}
    \end{subfigure}
    \hfill
    \begin{subfigure}{\linewidth}
        \centering        \includegraphics[width=0.5\linewidth]{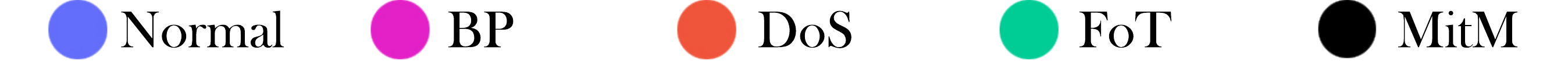}
    \end{subfigure}
    \hfill
    \caption{The visualization of the dataset.}
    \label{fig:dataset_visu}
    \vspace*{-0.5cm}
\end{figure}

We visualize the dataset as in Fig.~\ref{fig:dataset_visu}. We first use dimensionality reduction with t-distributed stochastic neighbourhood embedding (t-SNE)~\cite{tsne} to project the 21 features into three principal components or dimensions. Then, we plot the resulting 3D data points and colour code them according to their classes (normal behaviour or four types of attacks). Two important results can be noted. First, the spread and overlap of attack data points show that different attacks share some common features, which can make it harder for the CCD system to differentiate attack classes. Second, there is a clear variance between the 3D points of the same attack, reflecting the diversity in attack patterns and complicating attack detection and labeling by the CCD system.

\subsection{CCD System Setup}
We maintain a consistent configuration for the deep learning models across all clusters. For a fair comparison, we employ various types of deep learning networks, each configured with the same input layer, two hidden layers (each containing 128 neurons), and an output layer. The models operate with a batch size of 1024 and a learning rate of 0.0046. A collaborative learning mechanism facilitates collaborative learning among the different deep learning models in various clusters. This mechanism is implemented on an Intel computer with 8 CPUs running at 2592 MHz and 32 GB of RAM. Our collaborative learning system includes 100 clusters, each containing 1,470 samples for training. The specific deep learning model used can vary based on each test case. After the training process, a global model is generated, and a testing dataset comprising 63,000 samples is employed to evaluate the model's attack detection accuracy.

\subsection{Evaluation Method}
The confusion matrix~\cite{confusion_matrix1} is a ubiquitous method to evaluate the performance of many machine learning and deep learning systems in terms of accuracy, precision, and recall. We denote ``True Positive'',  ``True Negative'', ``False Positive'' and ``False Negative'' as TP, TN, FP and FN, respectively. The accuracy of a machine learning system can be calculated as follows:
\begin{equation}
\begin{aligned}
\label{eqn12}
	\mbox{Accuracy} = \frac{\mbox{Number}~\mbox{of}~\mbox{correct}~\mbox{predictions}}{\mbox{Total}~\mbox{number}~\mbox{of}~\mbox{samples}}.
\end{aligned}
\end{equation}

We denote $M$ as the total number of classification classes at the output layer and $m \in M$ as the classification class of the output. The precision of a machine learning system can be calculated as follows:
\begin{equation}
\begin{aligned}
\label{eqn13}
	\mbox{Pre} = \sum_{m=1}^M\frac{\mbox{TP}_m}{\mbox{TP}_m+\mbox{FP}_m}.
\end{aligned}
\end{equation}

\begin{figure}[!t]
    \centering
    \includegraphics[width=\linewidth]{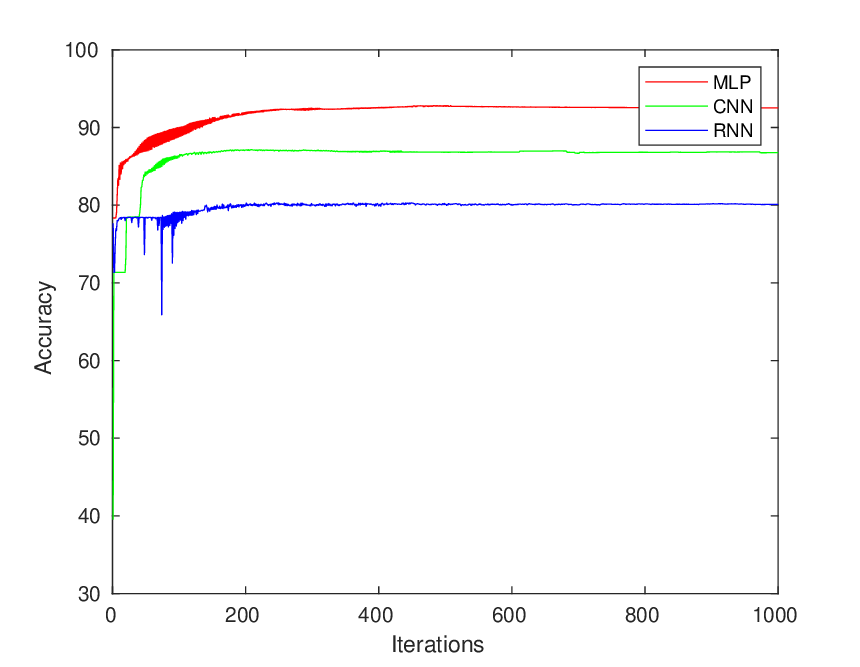}
    \caption{Securing CCD system with different DL models.}
    \label{fig:DL_models}
\end{figure}

\begin{figure*}[!t]
    \begin{subfigure}{0.34\linewidth}
        \centering
        \includegraphics[width=\linewidth]{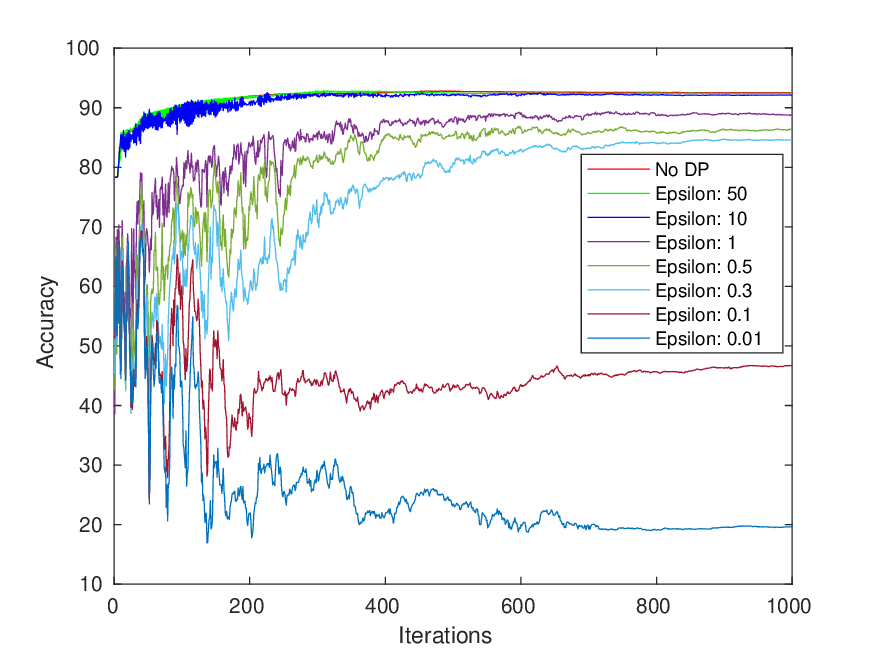}
        \caption{}
        \label{fig:Guassian_epsilon}
    \end{subfigure}
    \begin{subfigure}{0.34\linewidth}
        \centering
        \includegraphics[width=\linewidth]{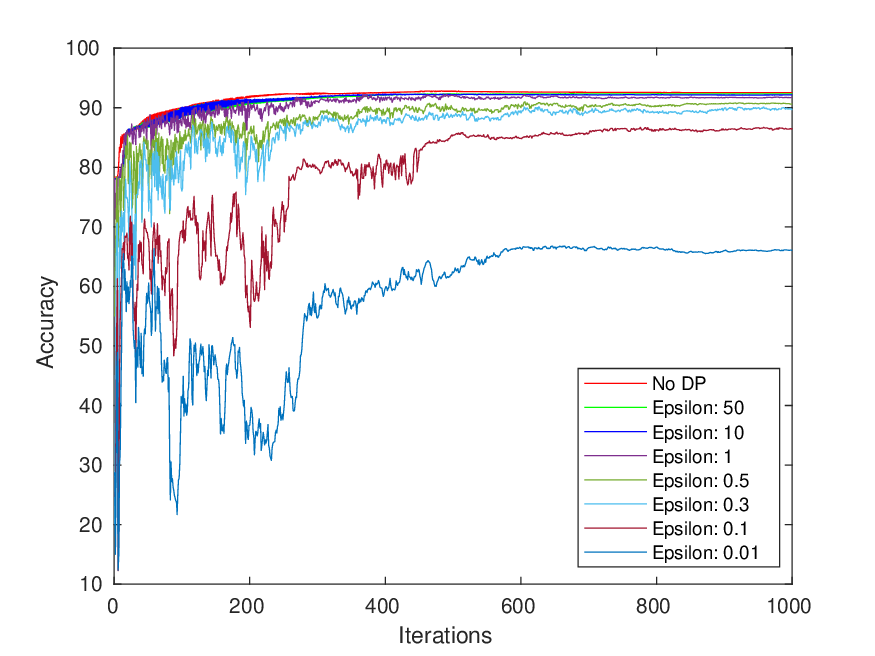}
        \caption{}
        \label{fig:Laplace_epsilon}
    \end{subfigure}
        \begin{subfigure}{0.34\linewidth}
        \centering
        \includegraphics[width=\linewidth]{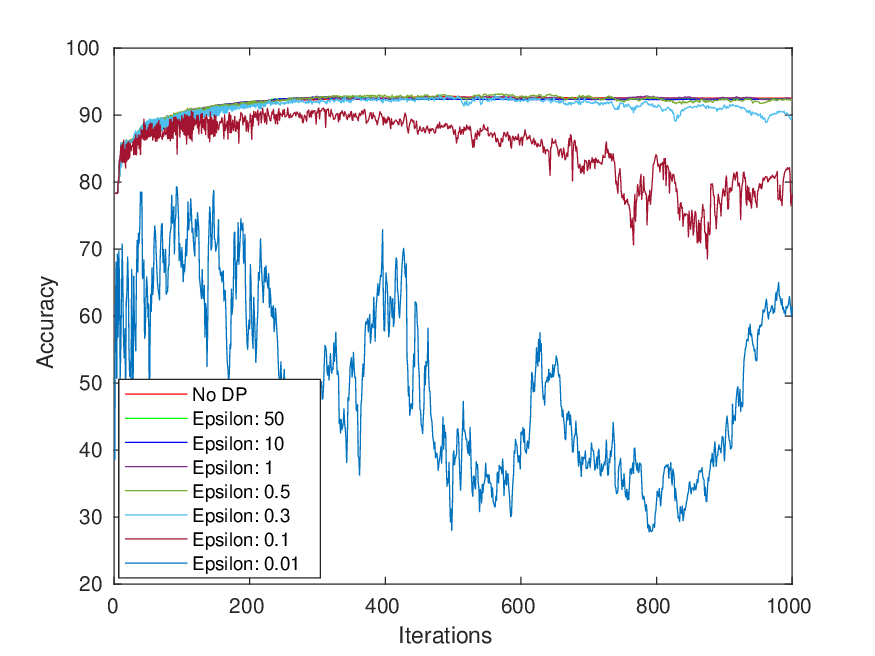}
        \caption{}
        \label{fig:MA_epsilon}
    \end{subfigure}
    \caption{Securing CCD system with different noises: (a) The Gaussian noises, (b) The Laplace noises, and (c) The Gaussian MA noises.}
    \label{fig:noise}
\end{figure*}

Similarly, the recall of the total system can be calculated as follows:
\begin{equation}
\begin{aligned}
\label{eqn14}
	\mbox{Re} = \sum_{m=1}^M\frac{\mbox{TP}_m}{\mbox{TP}_m+\mbox{FN}_m}.
\end{aligned}
\end{equation}

The $\mbox{Accuracy}$, $\mbox{Pre}$, $\mbox{Re}$ will be used to evaluate the performance of various models in the next section.

\section{Experiment Results and Performance Evaluation}
\label{sec:Performance_Evaluation}

\subsection{Performance of Various DL Models}

In this session, we evaluate the performance of different CCD setups without adding noise and three working clusters. We consider three ubiquitous types of machine learning models, including multilayer perceptron (MLP) networks, convolutional neural networks (CNN), and recurrent neural networks (RNN). Fig.~\ref{fig:DL_models} shows the performance in accuracy of these models. In Fig.~\ref{fig:DL_models}, we can see that after large fluctuations at the beginning, the accuracy of all models converged after about 1000 iterations. The MLP model achieves the highest accuracy and stability at around 92.5\% in comparison with those of CNN and RNN at 86.7\% and 80.1\%, respectively. These results demonstrate that in these experiments, the MLP model detects attacks better than CNN and RNN in the blockchain network. Based on this result, we use MLP in all clusters in the next section to consider the effects of adding noise in the CCD system.

\subsection{Impacts of Adding Noises}

In this section, we use the MLP model for CCD in each cluster. We study various types of noises, including Gaussian noises, Laplace noises, and MA noises with different $\epsilon$ (i.e., $0.01$, $0.1$, $0.3$, $0.5$, $1$, $10$, $50$, no noise ($\epsilon = \infty$)), to the model to evaluate the results. 

Fig.~\ref{fig:noise} provides the results of the experiments. In these experiments, we fix $\delta$ at $10^{-5}$ and the number of clusters at $3$. Fig.~\ref{fig:Guassian_epsilon} shows the effects of Gaussian noise with different cases of $\epsilon$. In the case of without noise ($\epsilon = \infty$), the MLP model of the CCD converges after $300$ iterations and gets an accuracy up to 92.5\%. This statement is also demonstrated by the accuracy of $\epsilon = 50$, which is nearly the same as the case of without noise. As shown in Fig.~\ref{fig:Guassian_epsilon}, when $\epsilon$ reduces, more Gaussian noises are added to the model, and the accuracy curves need more iterations to converge. In detail, while the accuracy curves with $\epsilon = 10$ and $\epsilon = 50$ need about 300 iterations for convergence, the others with lower $\epsilon$ converge after approximately 700 iterations. In addition, we can see in Fig.~\ref{fig:Guassian_epsilon} that after large fluctuations at the beginning, all the accuracy curves converge. After 1,000 iterations, the accuracy of each case reduces proportionally to the reduction of the $\epsilon$, especially with $\epsilon = 0.1$ and $\epsilon = 0.01$.

We can observe the same trend with Laplace noise in Fig.~\ref{fig:Laplace_epsilon} where the accuracy curves fluctuate at the beginning of $500$ iterations. After 1,000 iterations, the accuracy values decrease in proportion to the reduction of the value of $\epsilon$. In addition, compared to Fig.~\ref{fig:Guassian_epsilon} in cases of $\epsilon=0.01$ and $\epsilon=0.1$, the accuracies with Laplace noise are higher than those with Gaussian noise. 
This can demonstrate that the Laplace noise has less effect on the accuracy of cyberattack detection in blockchain networks than the Gaussian noise.
In Fig.~\ref{fig:MA_epsilon}, with MA noise, after 1,000 iterations, the accuracy curves are converged and nearly the same in the cases of no noise with $\epsilon = 50$, $\epsilon = 10$, and $\epsilon = 1$. However, we can observe a slight fluctuation of accuracy in the case of $\epsilon = 0.5$ around 92\% after 1,000 iterations. The fluctuations slightly increase with $\epsilon = 0.3$ but significantly increase when the $\epsilon$ reduces to $0.1$ and $0.01$. 
These results demonstrate that Gaussian and MA noises have stronger effects on the CCD models than the Laplace noise. Due to these results, we recognize that with $\epsilon = 0.5$ and $\epsilon = 0.3$, the added noises remarkably affect the CCD system while still keeping acceptable accuracy after 1,000 iterations. Thus, in the following section, we will evaluate how varying the number of clusters impacts the accuracy in attack detection across different noise levels, specifically for $\epsilon = 0.5$ and $\epsilon = 0.3$. This analysis will provide crucial insights into the robustness of our detection approach under varying conditions.

\subsection{Impacts of Number of Participating Clusters}
\begin{figure*}[!t]
    \centering
    \begin{subfigure}{0.45\linewidth}
        \includegraphics[width=\linewidth]{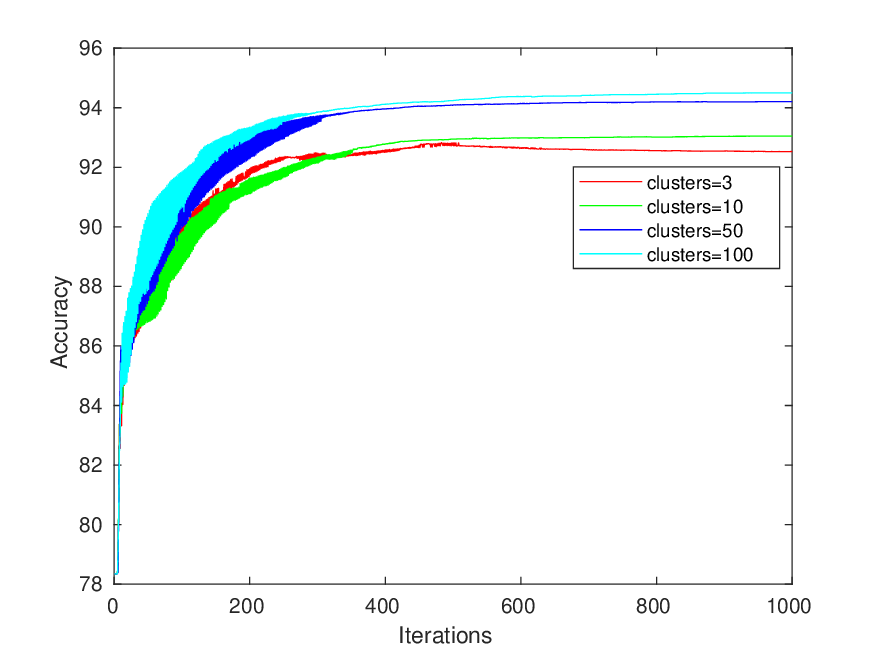}
        \caption{}
        \label{fig:nonoise_clusters}
    \end{subfigure}
    \hfill
    \begin{subfigure}{0.45\linewidth}
        \includegraphics[width=\linewidth]{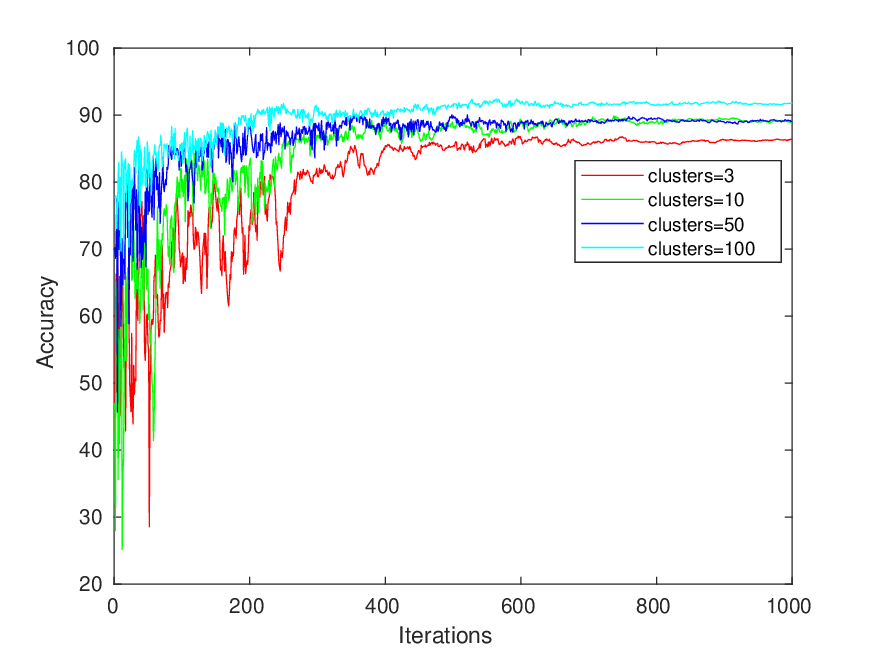}
        \caption{}
        \label{fig:Gaussian_clusters}
    \end{subfigure}
    \hfill
    \begin{subfigure}{0.45\linewidth}
        \includegraphics[width=\linewidth]{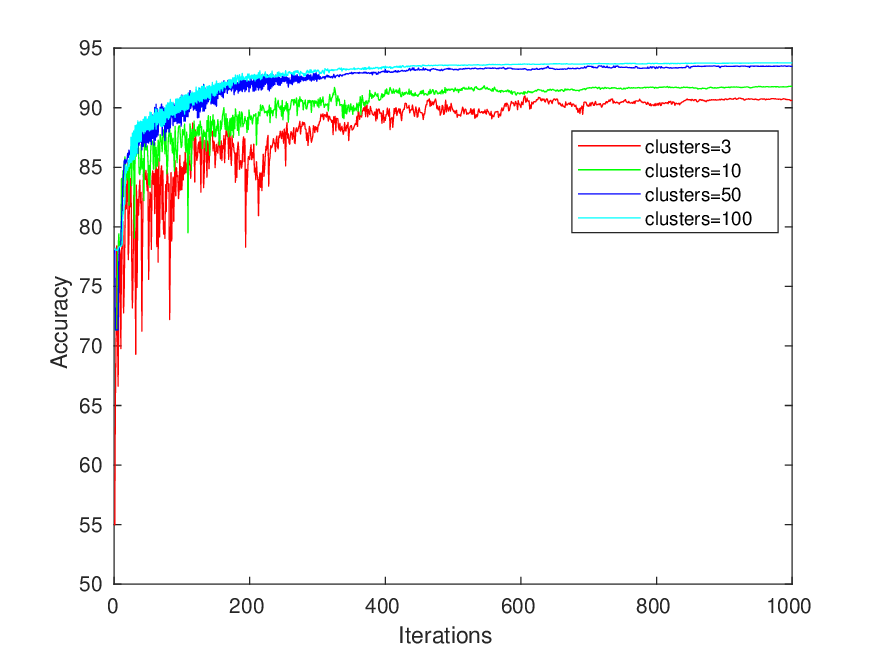}
        \caption{}
        \label{fig:Laplace_clusters}
    \end{subfigure}
    \hfill
    \begin{subfigure}{0.45\linewidth}
        \includegraphics[width=\linewidth]{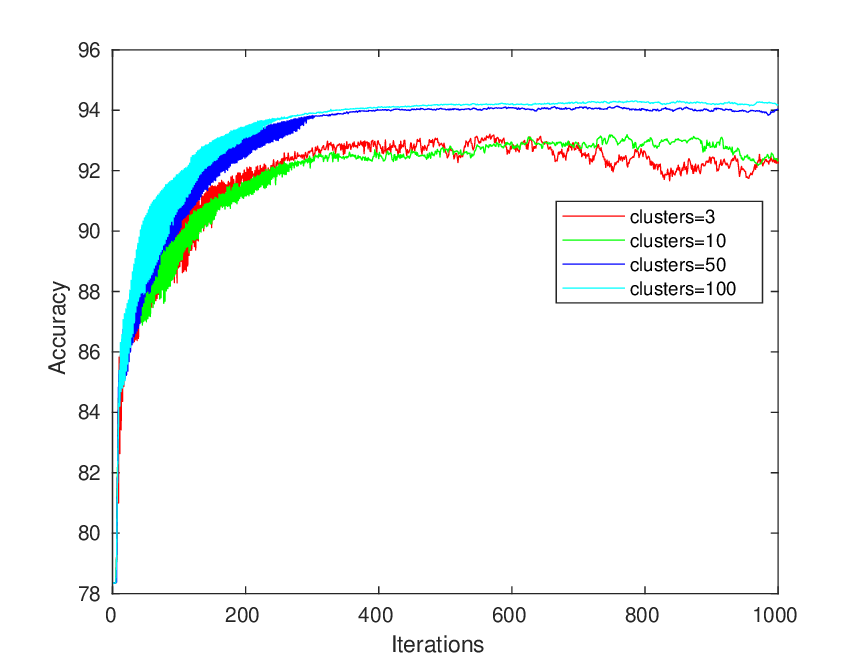}
        \caption{}
        \label{fig:MA_clusters}
    \end{subfigure}
    \hfill
    \caption{Securing CCD results with various noises in different cases of clusters with $\epsilon = 0.5$: (a) Different clusters without noise, (b) Gaussian noise with different clusters, (c) Laplace noise with different clusters, and (d) MA noise with different clusters.}
    \label{fig:clusters}
    \vspace*{-0.5cm}
\end{figure*}

\begin{figure*}[!t]
    \begin{subfigure}{0.34\linewidth}
        \centering
        \includegraphics[width=\linewidth]{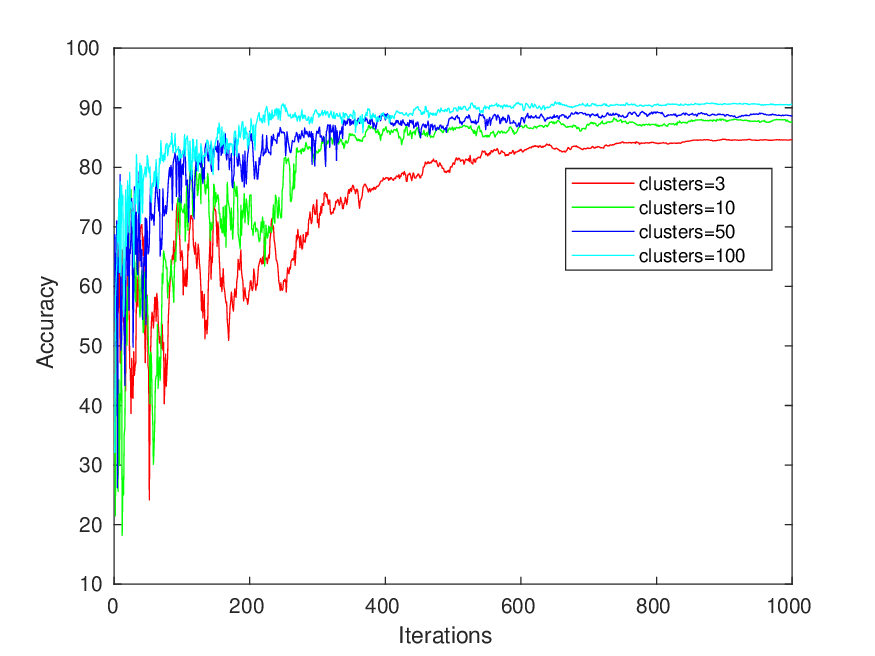}
        \caption{}
        \label{fig:FL_Gaussian_0.3}
    \end{subfigure}
    \begin{subfigure}{0.34\linewidth}
        \centering
        \includegraphics[width=\linewidth]{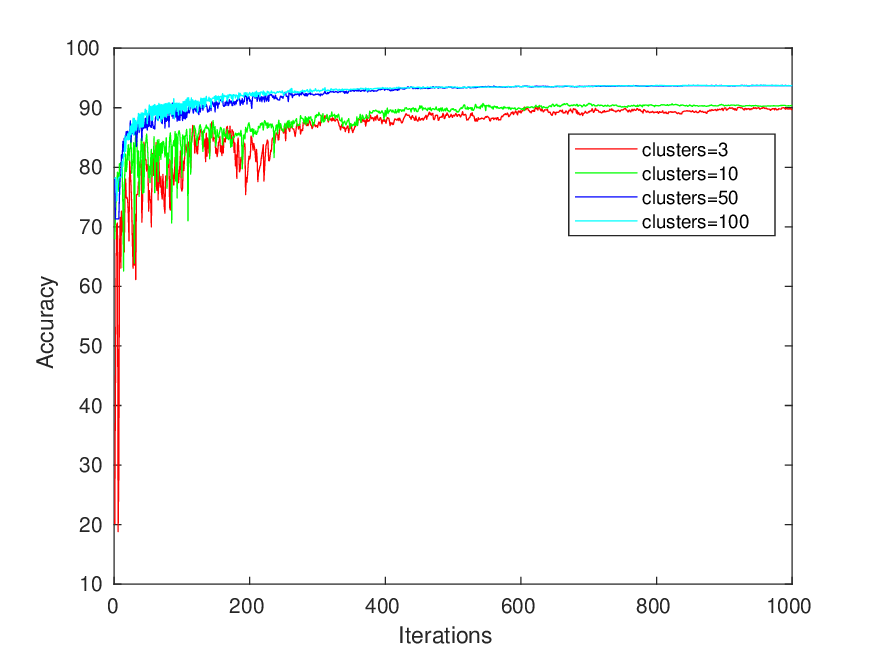}
        \caption{}
        \label{fig:FL_Laplace_0.3}
    \end{subfigure}
        \begin{subfigure}{0.34\linewidth}
        \centering
        \includegraphics[width=\linewidth]{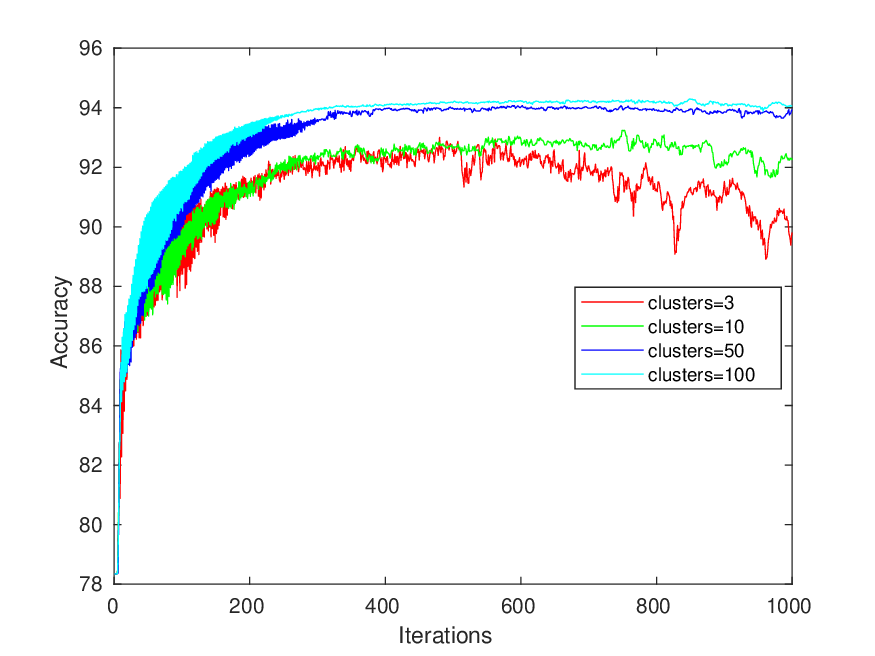}
        \caption{}
        \label{fig:FL_MA_0.3}
    \end{subfigure}
    \caption{Securing CCD results with various noises in different cases of clusters with $\epsilon = 0.3$: (a) Gaussian noise with different clusters, (c) Laplace noise with different clusters, and (d) MA noise with different clusters.}
    \label{fig:othernoise}
\end{figure*}

\begin{figure*}[!t]
    \centering
    \begin{subfigure}{0.45\linewidth}
        \includegraphics[width=\linewidth]{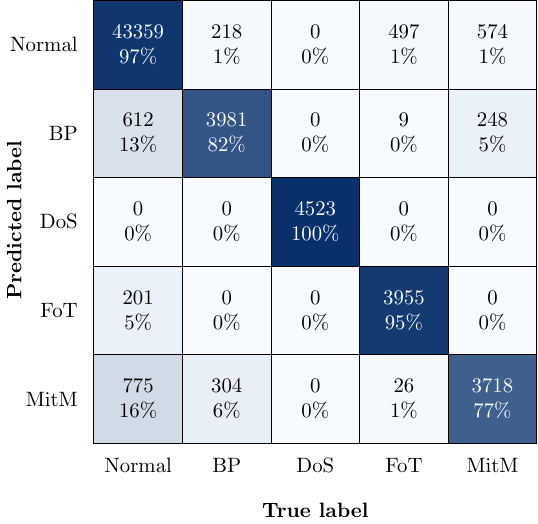}
        \caption{}
        \label{fig:nonoise_cfm}
    \end{subfigure}
    \hfill
    \begin{subfigure}{0.45\linewidth}
        \includegraphics[width=\linewidth]{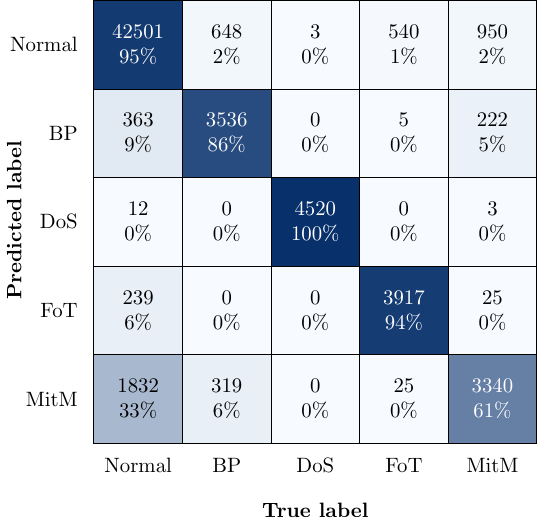}
        \caption{}
        \label{fig:Gaussian_cfm}
    \end{subfigure}
    \hfill
    \begin{subfigure}{0.45\linewidth}
        \includegraphics[width=\linewidth]{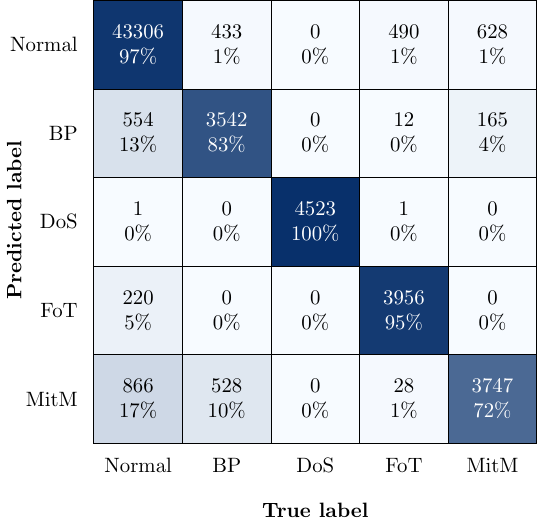}
        \caption{}
        \label{fig:Laplace_cfm}
    \end{subfigure}
    \hfill
    \begin{subfigure}{0.45\linewidth}
        \includegraphics[width=\linewidth]{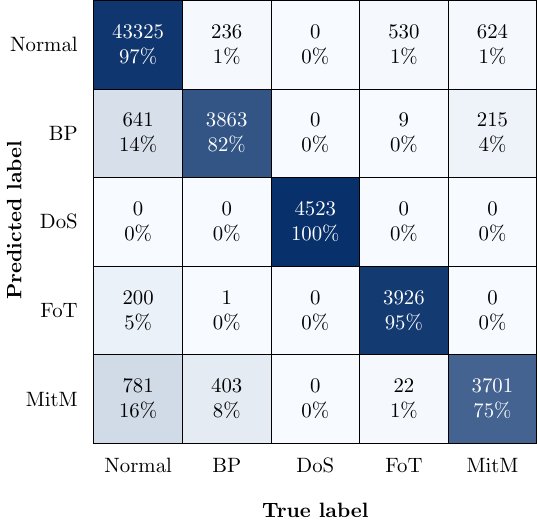}
        \caption{}
        \label{fig:MA_cfm}
    \end{subfigure}
    \hfill
    \caption{The confusion matrix for each class detection in a cluster with Epsilon=0.5: (a) Without noise, (b) Gaussian noise, (c) Laplace noise, and (d) MA noise}
    \label{fig:cfm}
    \vspace*{-0.5cm}
\end{figure*}

In this section, we consider noise parameters of Gaussian, Laplace and MA with $\epsilon = 0.5$, $\epsilon = 0.3$ and $\delta = 10^{-5}$. We then vary the number of clusters to measure the effects of noises as the number of samples increases. 

Fig.~\ref{fig:clusters} provides the results of these simulations in the case of $\epsilon = 0.5$ and $\delta = 10^{-5}$. In particular, Fig.~\ref{fig:nonoise_clusters} shows the accuracy curves without noise with 3 clusters, 10 clusters, 50 clusters, and 100 clusters. In this figure, we can observe that when the number of samples for training increases significantly from 3 clusters to 100 clusters, all of the curves converge after about 400 iterations, and the accuracy increases proportionally to the number of clusters. This is because the more clusters there are, the more samples will be added to the training model. Fig.~\ref{fig:Gaussian_clusters} and Fig.~\ref{fig:Laplace_clusters} illustrate the accuracy curves after adding Gaussian noise and Laplace noise, respectively. It is observed that after large fluctuations at the first 400 iterations, the accuracy curves of these figures converge the same as those in Fig.~\ref{fig:nonoise_clusters}. However, due to the added noises, the accuracy after 1,000 iterations reduces considerably, especially with Gaussian noise from $92.5$\% in the case of no noise with 3 clusters to $86.4$\% in the case of Gaussian noise with 3 clusters.
Fig.~\ref{fig:MA_clusters} provides information about the effects of MA noise on the CCD system. We can see in this figure that the MA noises have strong effects in the case of 3 clusters and 10 clusters. The accuracy curves in these cases still have a number of fluctuations around $92$\% at the end of 1000 iterations. However, it is more stable and converged at about $94$\% when the number of clusters increases to 50 and 100. It is important to note that across all scenarios with added noise, the accuracy curves for attack detection in the collaborative learning model consistently converge and improve as the number of clusters increases. This trend underscores the model's resilience and its enhanced detection capability with a more refined clustering approach.

Fig.~\ref{fig:othernoise} describes the accuracy curves of the cyberattack detection model in the blockchain network with $\epsilon = 0.3$ and $\delta = 10^{-5}$ with Gaussian, Laplace and MA noises. In these figures, the added noises to the cyberattack detection model are stronger than those in Fig.~\ref{fig:clusters}. Even though the added Gaussian and Laplace noises in Fig.~\ref{fig:FL_Gaussian_0.3} and Fig.~\ref{fig:FL_Laplace_0.3} have the same trend as in Fig.~\ref{fig:Gaussian_clusters} and Fig.~\ref{fig:Laplace_clusters}, the accuracy of cyberattack detection slightly reduces, e.g., from $91.7$\% to $90.58$\% in case of Gaussian noise with 100 clusters and from $93.75$\% to $93.738$\% in case of Laplace noise with 100 clusters. In Fig.~\ref{fig:FL_MA_0.3}, we can observe significant fluctuations of Moment Accountant in the case of 3 clusters at the end of 1,000 iterations. The fluctuations decrease when the number of clusters increases from 3 to 10. These fluctuations are nearly removed, and the accuracy curves are stable and converged in the case of 50 and 100 clusters. From Fig.~\ref{fig:othernoise}, we can see that with the same parameters (i.e., $\epsilon$ and $\delta$), the MA noise has a stronger effect on the cyberattack detection model in the cases of 3 and 10 clusters.


Table~\ref{tab:perform_results} describes the detailed results of the performance in accuracy, precision, and recall in the cases of 3 clusters, 10 clusters, 50 clusters and 100 clusters and $\epsilon=0.5$ with normal CCD system and different types of noises. Because of the fluctuations at the end of 1,000 iterations with MA noise, the accuracy, precision, and recall are calculated by taking average values of the last 100 iterations (i.e., the average value from iteration 900 to 1000). First, it is observed that the accuracy increases proportionally to the number of clusters. However, in general, the accuracy with added noises is smaller than the accuracy without noise in all the cases as shown in Table~\ref{tab:perform_results}. Second, the accuracy of Moment Accountant with 100 clusters achieves the best value at $94.22$\% in all cases with added noises. Besides, the lowest accuracy is at $86.26$\% belonging to Gaussian noise with 3 clusters. The Laplace noise provides average accuracy compared to Gaussian noise and MA noise. Fig.~\ref{fig:cfm} provides the confusion matrix for no noise, Gaussian noise, Laplace noise, and MA noise in the case of 100 clusters with $\epsilon=0.5$. In this figure, we can look deeply into the accuracy of each classification class. We can see that, in this case, the added noise has a strong effect on the MitM attacks that leads to the reduction of the total accuracy of the whole system. However, there is a slight increase in the accuracy of attack detection on BP attacks at 86\% with Gaussian noise in comparison with 82\% without noise. The reason is that, in this case, the DP parameters play a role as a regularization factor for the deep learning model to improve the accuracy of attack detection.



\subsection{CCD Processing Time}

\begin{table*}[!t]
    \centering
    \caption{The performance results of the simulations.}
    \label{tab:perform_results}
    \resizebox{\linewidth}{!} {
\begin{tabular}{|c|cccc|cccc|cccc|cccc|}
\hline
\multicolumn{1}{|l|}{}                            & \multicolumn{4}{c|}{\textbf{No noise}}                                                                               & \multicolumn{4}{c|}{\textbf{Gaussian noise}}                                                                         & \multicolumn{4}{c|}{\textbf{Laplace noise}}                                                                          & \multicolumn{4}{c|}{\textbf{MA noise}}                                                                \\ \hline
\multicolumn{1}{|l|}{\textbf{Number of clusters}} & \multicolumn{1}{c|}{\textbf{3}} & \multicolumn{1}{c|}{\textbf{10}} & \multicolumn{1}{c|}{\textbf{50}} & \textbf{100} & \multicolumn{1}{c|}{\textbf{3}} & \multicolumn{1}{c|}{\textbf{10}} & \multicolumn{1}{c|}{\textbf{50}} & \textbf{100} & \multicolumn{1}{c|}{\textbf{3}} & \multicolumn{1}{c|}{\textbf{10}} & \multicolumn{1}{c|}{\textbf{50}} & \textbf{100} & \multicolumn{1}{c|}{\textbf{3}} & \multicolumn{1}{c|}{\textbf{10}} & \multicolumn{1}{c|}{\textbf{50}} & \textbf{100} \\ \hline
\textbf{Accuracy}                                 & \multicolumn{1}{c|}{92.53}      & \multicolumn{1}{c|}{93.04}       & \multicolumn{1}{c|}{94.2}        & 94.5         & \multicolumn{1}{c|}{86.26}      & \multicolumn{1}{c|}{89.16}       & \multicolumn{1}{c|}{89.09}       & 91.71        & \multicolumn{1}{c|}{90.72}      & \multicolumn{1}{c|}{91.74}       & \multicolumn{1}{c|}{93.48}       & 93.75        & \multicolumn{1}{c|}{92.97}      & \multicolumn{1}{c|}{92.55}       & \multicolumn{1}{c|}{93.98}       & 94.22        \\ \hline
\textbf{Precision}                                & \multicolumn{1}{c|}{86.75}      & \multicolumn{1}{c|}{87.7}        & \multicolumn{1}{c|}{90.14}       & 90.29        & \multicolumn{1}{c|}{76.5}       & \multicolumn{1}{c|}{80.78}       & \multicolumn{1}{c|}{82.69}       & 86.96        & \multicolumn{1}{c|}{84.02}      & \multicolumn{1}{c|}{84.88}       & \multicolumn{1}{c|}{88.45}       & 89.28        & \multicolumn{1}{c|}{85.89}      & \multicolumn{1}{c|}{86.89}       & \multicolumn{1}{c|}{89.67}       & 89.89        \\ \hline
\textbf{Recall}                                   & \multicolumn{1}{c|}{87.87}      & \multicolumn{1}{c|}{87.84}       & \multicolumn{1}{c|}{90.02}       & 90.98        & \multicolumn{1}{c|}{76.21}      & \multicolumn{1}{c|}{81.75}       & \multicolumn{1}{c|}{88.03}       & 86.54        & \multicolumn{1}{c|}{83.82}      & \multicolumn{1}{c|}{86.84}       & \multicolumn{1}{c|}{89.71}       & 89.14        & \multicolumn{1}{c|}{87.36}      & \multicolumn{1}{c|}{87}          & \multicolumn{1}{c|}{89.91}       & 90.27        \\ \hline
\end{tabular}
}
\end{table*}

\begin{figure}[!t]
    \centering
    \includegraphics[width=\linewidth]{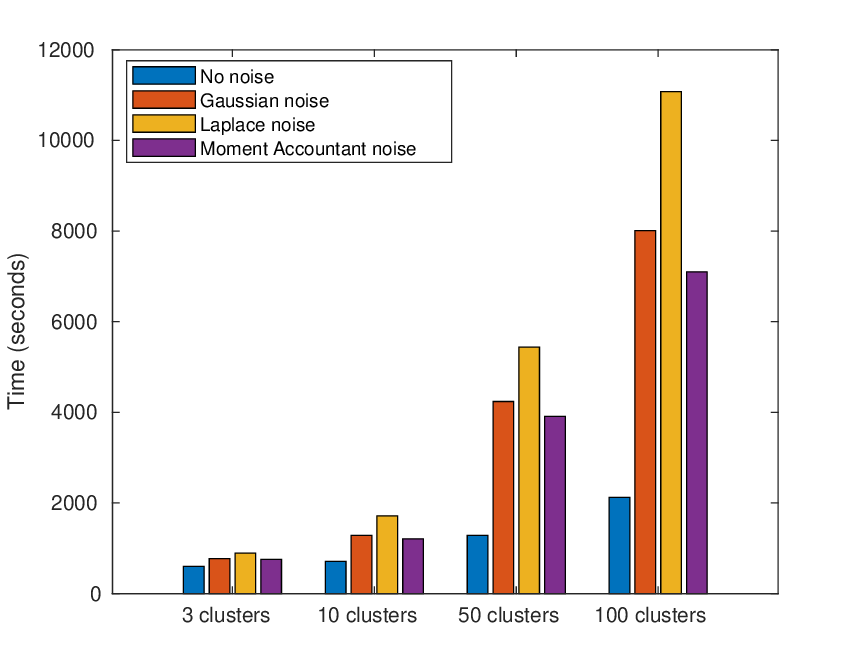}
    \caption{The computational time of CCD systems.}
    \label{fig:time}
\end{figure}

Fig.~\ref{fig:time} shows the computational time of each model. In Fig.~\ref{fig:time}, we can see that even though all models converge after 1,000 iterations, the computational time is different with the same number of clusters. The added noise systems also need more processing than the original without noise system. In addition, the computational time of 100 clusters is much longer than others. The reason is that the CCD systems with 100 clusters have more samples than those with other clusters. Thus, they need more time to aggregate data from all clusters in each iteration than others. In addition, we can see in this figure that the Laplace noise needs more computational time in comparison with Gaussian and MA noise. These results reveal that there are various trade-offs among the types of added noises, the number of clusters, accuracy and the convergence time. For example, the MA noise has the best performance in accuracy with both $\epsilon=0.5$ and $\epsilon=0.3$ in the cases of 50 and 100 clusters and lower computational time. However, it is unstable in the cases of 3 and 10 clusters with $\epsilon=0.3$ as can be seen in Fig.~\ref{fig:FL_MA_0.3}. The Laplace noise can provide high performance in the cases of 3 and 10 clusters with $\epsilon=0.3$ as in Fig.~\ref{fig:FL_Laplace_0.3}. However, this approach requires more computational time compared to other noise types. While Gaussian noise offers a more consistent computational time, it yields lower accuracy after convergence in both $\epsilon=0.5$ and $\epsilon=0.3$ scenarios with 3 and 10 clusters, particularly when compared to Laplace noise.

\section{Conclusion}\label{sec:conclusion}
In this paper, we have introduced an innovative privacy-preserving solution for Collaborative Cyberattack Detection (CCD) in blockchain networks, addressing a critical challenge in modern cybersecurity. By incorporating noise into trained models before collaborative learning, we have demonstrated a robust approach to safeguarding sensitive data while maintaining the efficacy of attack detection. Our extensive simulations evaluated the impact of different noise types, i.e., Gaussian, Laplace, and Moment Accountant, on key performance metrics such as accuracy, convergence time, and overall runtime across varying numbers of clusters. The results underscore the complex interplay between noise addition and system performance, revealing that while noise is essential for privacy, its effects must be carefully managed to avoid compromising detection capabilities. This work not only advances our understanding of the balance between data privacy and performance in CCD systems but also offers practical guidelines for optimizing these parameters in diverse environments. In future work, we will continue to explore innovative methods to enhance the protection of federated learning-based models, ensuring that local data remains secure in increasingly complex network environments. This research lays a strong foundation for future advancements in the field, paving the way for more secure, reliable, and scalable blockchain-based data-sharing systems.

\bibliographystyle{IEEEtran}
\bibliography{library}

\end{document}